# Pandora: A Cyber Range Environment for the Safe Testing and Deployment of Autonomous Cyber Attack Tools

Hetong Jiang, Taejun Choi and Ryan K. L. Ko

The University of Queensland, St. Lucia, Brisbane, Queensland, Australia
{hetong.jiang,taejun.choi,ryan.ko}@uq.edu.au

**Abstract.** Cybersecurity tools are increasingly automated with artificial intelligent (AI) capabilities to match the exponential scale of attacks, compensate for the relatively slower rate of training new cybersecurity talents, and improve of the accuracy and performance of both tools and users. However, the safe and appropriate usage of autonomous cyber attack tools – especially at the development stages for these cyber attack tools – is still largely an unaddressed gap. Our survey of current literature and tools showed that most of the existing cyber range designs are mostly using manual tools and have not considered augmenting automated tools or the potential security issues caused by the tools. In other words, there is still room for a novel cyber range design which allow security researchers to safely deploy autonomous tools and perform automated tool testing if needed. In this paper, we introduce Pandora, a safe testing environment which allows security researchers and cyber range users to perform experiments on automated cyber attack tools that may have strong potential of usage and at the same time, a strong potential for risks. Unlike existing testbeds and cyber ranges which have direct compatibility with enterprise computer systems and the potential for risk propagation across the enterprise network, our test system is intentionally designed to be incompatible with enterprise real-world computing systems to reduce the risk of attack propagation into actual infrastructure. Our design also provides a tool to convert in-development automated cyber attack tools into to executable test binaries for validation and usage realistic enterprise system environments if required. Our experiments tested automated attack tools on our proposed system to validate the usability of our proposed environment. Our experiments also proved the safety of our environment by compatibility testing using simple malicious code.

## 1    Introduction

The increasing rate and global nature [1] of cyber attacks and the recent trends towards cyber autonomy [2] have resulted in questions about the scalability of current cybersecurity approaches and the relatively-long time required to train cyber professionals in comparison to the emergence of new cyber threats [3]. In an attempt to overcome this asymmetric attack and response rates, we are witnessing an increasing demand for cyber autonomy from the cyber-security industry. There is also an inability to train cybersecurity professionals quick enough to be able to address the increasing rate of cyber threats.

Donevski and Zia [4] stated that the current lack of cybersecurity professionals is one of the reasons the cybersecurity industry has struggled in addressing this scale problem. This is also widely termed as the 'cyber skills gap'. According to Smith [5], while it is a good effort, re-education may not a practical solution, as the rate of cyber threats is accelerating too fast for existing professionals to handle – overwhelming many cybersecurity professionals in the process. An augmentation of automation tools to automate most tasks while freeing up time and resources for cybersecurity professionals to make decisions aided by cyber automation promises to meet the exponentially-growing demands for cybersecurity, risk assessments and compliance [6]. Cyber autonomy also allows the cybersecurity industry address scale and new security issues introduced by cloud computing [7-9] and other emerging technologies.

Besides addressing the cyber skills gap and the accelerating rate of cyber threats, it is also important to be able to evaluate and test the usefulness of autonomous cyber attack tools often used in increasingly-automated red-teaming environments or security assessments. Having a technique or platform to test these cyber autonomy tools will facilitate more effective tool evaluation, especially before they are actually deployed into actual enterprise environment. Researchers have also found automated tools with AI algorithms to be cost effective solutions [10]. Furthermore, a diverse set of test areas for system hardening could be conducted by automated tools such as detecting SQL injection vulnerabilities [11], penetrating testing [12], man-in-the-middle attacks[13] and finding exploitable bugs in binaries [14].

Despite the usefulness of automated cyber attack tools, prototyped versions of the tools in their initial stages of development could cause unexpected results. Kaloudi and Li's survey paper [15] and the DARPA Cyber Grand Challenge (CGC) [16] described and demonstrated fully-automated AI-based automated attacks could evade detection measures respectively. The risks around deploying these tools before proper testing are evident. Without first testing within a proper testing environment, a prototyped automated attack tool may cause critical damage on real-world infrastructures. In fact, a prominent example is the Morris worm in 1988 [17], which was initially created to understand the size of the Internet at that time, but unintentionally rendered about 10% of the systems on the Internet unusable.

Despite possible threats caused by the unintended situations, the automation of exploitation tools is a popular topic for cybersecurity researchers. Web-based attack tools such as vulnerability scanning [1, 18], SQL Injection [11, 19] and Cross-Site Scripting [19] are the most commonly seen topics for network automation tools. Consequently, testing environments for network-based tools are commonly proposed.

In the case of binary exploitations, at the time of writing, there are few existing testing environments that are designed for automated binary exploit tools. Researchers usually utilize virtualization technologies to create test environments for binary experiments. However, this does not guarantee a safe environment. For example, vulnerabilities such as Venom [20] (CVE-2015-3456) and VMware Fusion's vulnerability (VMSA-2015-0004) (CVE-2015-2337) [21] allow malicious code to penetrate the barrier between VMs and Host, thereby causing damage to host systems.

Even with proper security measurements, some attack tools, especially automated tools with advanced technologies such as machine learning could be unpredictable [22]. Conversely, machine learning techniques could also help malware avoid detection. There is also no guarantee with regards to the outputs generated during the testing – putting the host system in increased risks. Therefore, a secure testing environment is required to prevent unpredictable propagation of risks into connected enterprise environments – achieving the true requirements of a 'sandboxed' environment.

To mitigate the above gaps and provide a solution for cyber ranges and test environments, we propose Pandora, a secure cyber range design for automated cybersecurity tools testing. The proposed design allows automated testing of any types of exploits without threatening the real-world infrastructures linked to the cyber range. Our approach also caters to the requirements of deployment fully-automated tools within the cyber range – preparing the users for an increasingly autonomous cyber environment.

In order to prove the functionality of the proposed cyber range, a vulnerable binary in our cyber range has been successfully and automatically exploited by a binary exploit tool. This was formed via combining a number of known cybersecurity tools, to represent an automated attack.

We carried out a comparison of simple malicious code in our suggested environment and a Linux Ubuntu 18.04 LTS Linux system to confirm the isolation of automated attack tools from real-world infrastructures connected to the cyber range. The code worked and achieved our goals within our isolated environment, but it does not work for any generic Linux system – proving the non-portability of the test from our environment to typical enterprise Linux environments.

The main contributions of our research are:

— We propose a cyber range design to mitigate potential risks of automated cybersecurity tools in their early-stage development. The design could be implemented using common off-the-shelf tools.
— We verified the proper working of our implementation by executing an automated tool in our system.
— We confirm the isolation of the designed cyber range from real-world infrastructures – achieving non-propagation of potential unknown risks.

Our paper is organized as follows. Section 2 discuss the background and related work. We provide the design of our proposed cyber range in Section 3 and in Section 4, we discuss case studies including an overview of hardware systems and software used. Result analyses of case studies including verification for incompatibility of the implementation are discussed in Section 5. We then conclude and propose future work in Section 6.

## 2  Related Work

### 2.1  Background

As described in [23], a cyber range is an environment that provides a realistic environment suitable for conducting 'live fire' type of exercises which train computer network operators for cyber defense, and support experimentation and testing via a combination of cybersecurity products. Typically, the architecture of a cyber range could be split into sub-components. For example, as shown in Figure 1, Yamin *et al.* [24] suggested eight components comprising a cyber range. As long as the setup is correct and there are no misuse of tools, a cyber range environment allows anyone who wishes to have cybersecurity-related experiential training in cyber attack and defense to train within an environment with no direct risks on actual computer systems.

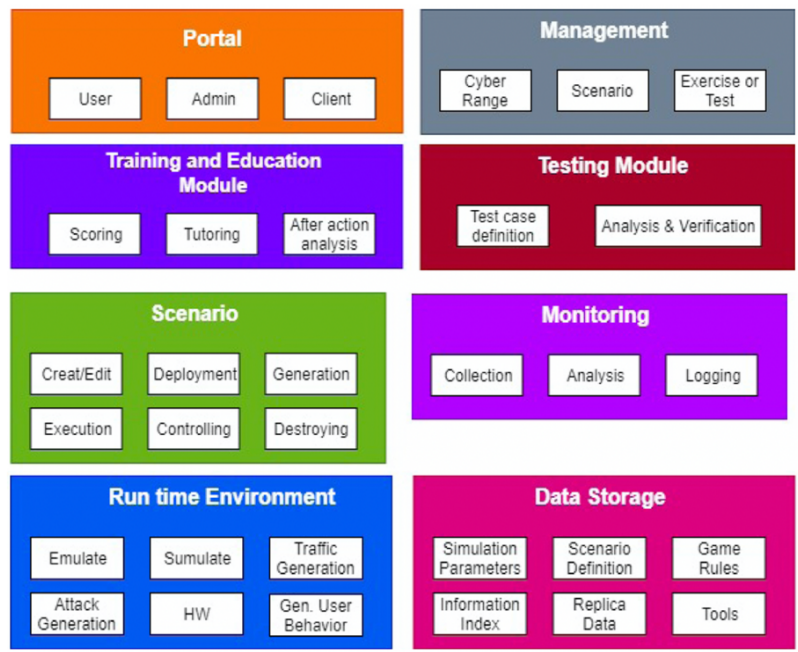

**Fig. 1.** Cyber range and security testbed functional architecture [1]

Considering the eight components of Figure 1, our research focuses on the 'Runtime Environment' which provides a safe test environment for cybersecurity researchers. The 'Runtime Environment' component is composed of the 'Emulator', 'Simulator', 'Attack Generation', 'HW' (i.e. hardware), 'Traffic Generation', and 'Gen. User Behavior' subcomponents. However, the 'Traffic Generation' and 'Gen. User Behavior' are out of scope for this paper which focuses on demonstrating test environments for automated binary exploitation cyber attack tools.

In the case of the subcomponent 'HW', real physical hardware are required to provide computing power. Based on the 'HW', an 'Emulate' subcomponent could mimic an operating system to create a simulated test environment. To create a simulated victim system, vulnerable applications or disclosed common vulnerabilities and exposures (CVE[25]) could be set up on the emulated operating system. With the aforementioned subcomponents, an attack application automated will be tested as a subcomponent 'Attack Generation' as shown in Figure 1.

Unlike many existing works that focusing on virtualization or containerization of test environments[24, 26-29], we concentrate on the security and simplicity of the testing environment. The reason for that many designs are focused on virtualization technologies is to increase the realism of the testing environment, thus, helping researchers to learn in an environment that may simulate a certain situation during a cybersecurity incident –making the testing as realistic as possible. However, as mentioned in Section 1, virtualization may not guarantee a threat free environment for its test infrastructure since vulnerabilities such as Venom [20] and VMware Fusion's vulnerability [21].

## 2.2 Related Research

While AI enhances the effectiveness of tools, it can also be used by malware. For example, Goosen [22] showed that, even with proper security measurements, some attack tools – especially automated tools with advanced technologies such as machine learning – could help malware avoid detection. On the other hand, we are seeing the emergence of other branches of AI, such as Amos-Binks' usage of automated planning (which is concerned with the realization of strategies or action sequence based on goals and constraints) for execution by intelligent agents [30].

In recent years, automated binary exploitation tools have emerged. One of the most prominent projects, Mayhem [14] by ForAllSecure has shown that it is possible to let a machine exploit binaries by itself using regular binary executables. While the current methods were not as good as human cybersecurity specialists with more creativity to invent new classes of attacks [2], the method has actually been realized into the computer, and computers generally spend less time examining binaries and are usually much more effective on a larger scale [2]. Some other automated approaches were able to search, discover, and patch vulnerabilities binaries. An example is the Cyber Reasoning System (CRS) Rubeus [31] by Raytheon during the DARPA CGC which uses machine learning and other AI technologies to help enhance its offensive abilities.

Over time, cyber ranges have been designed for different purposes. For example, the DETER Project [32] is one of the most prominent and earliest projects, and was designed for scientific and cybersecurity purposes. One of the essential stand-out characteristics of the DETER project was how realistic it was in terms of mimicking actual IT infrastructures.

Some other recent implementations are the cyber range for Cyber Defense Situation Awareness from Debatty and Mees [26], and the ones taking the approach for educational purposes such as the RC$^2$F framework[27] and the SEED Lab [28]. A few other cyber range designs focus on increasing number of features and efficiency of challenge maintenance such as Yamin *et al.* [24], and the one by Frank *et al.* [29].

## 3 Proposed Framework

The related work above has revealed opportunities for new cyber range designs to accommodate cyber autonomy safely and securely, and retain the comprehensiveness and usability such that new researchers or students who would like to test or create new cyber autonomy tools can safely test the early-stage deployment without any risk of propagating unpredictable results to a connected enterprise network.

As such, one of the primary reasons for creating Pandora as a cyber range for automated tool testing is to contain the potential propagation of undesired outcomes of some automated tools. Therefore, our proposed framework should be able to help researchers test dangerous automated tools within a safe and secure environment, and also simple enough to work and analyze with. This means students and newcomers could also create their automated tools for small projects and experiments.

### 3.1 Threat Modelling

To encapsulate our framework's resilience against common threats faced by current approaches listed above, we developed a threat model for analysis. Our threat model is based on security vulnerabilities related to virtualization and communication.

#### 3.1.1 Threat for Virtualization

As [20, 21] demonstrated that the barrier to segregate a virtual machine from its host machine is unsafe, we shall assume that an automated attack tool could affect its host machine beyond the test environment. In the case of a highly realistic virtual environment, the damages from tested tools could be severe depending on the level of emulation. Our threat model focus on protecting realistically set-up infrastructures from unexpected attacks.

#### 3.1.2 Threat for Communication

Vulnerabilities such as CVE-2018-10933 [33] demonstrated possibilities that might help the malicious clients get unauthorized access to other machines. In this vulnerability, the emulated network connection could be utilized as a path to allow a cybersecurity tool/weapon in development to access systems outside the testbed or cyber range. Other network security and communication vulnerabilities are already researched and hence, the protection of internal or external network for cyber range is out of scope for this paper.

## 3.2 Framework Design

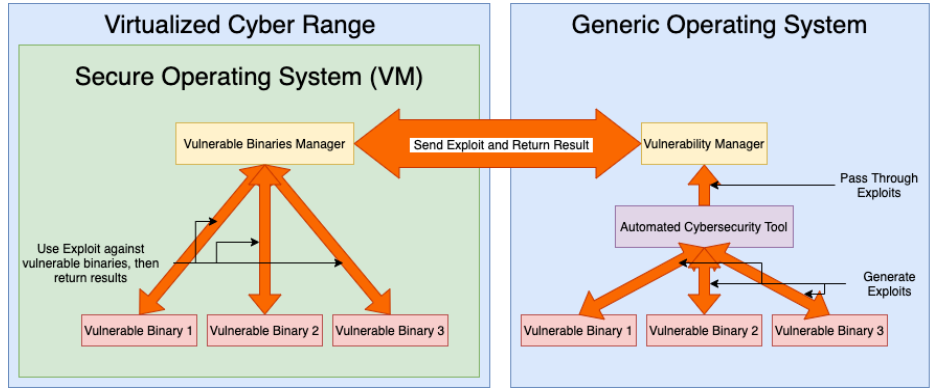

**Fig. 2.** Proposed Cyber Range Design for Cyber Autonomy

With our attack models described, we now propose a design of a cyber range focused on the "run time environment" illustrated previously in Figure 1, for testing automated cyber attack tools based on the modeled threats.

As shown in Figure 2, our proposed design could be divided into two main parts: 'Virtualized Cyber Range' and 'Generic Operating System' and their subcomponents. A brief description about the design is described below and details such as code snippets of the specific implementation can be found later in Section 5.

### 3.2.1 Virtualized Cyber Range

Our proposed cyber range is designed to be installed within a virtual machine, which should have a secure operating system which has some level of incompatibility with any generic operating system. As discussed earlier, the intentional incompatibility is introduced to reduce risks of unpredicted propagation of damage caused by tools to environments outside the test operating system. A vulnerable binary manager will be running for managing all exploit and analysis work.

─ *Secure Operating System (Secure OS)*

One of the ways that keep the cyber range secure at the operating system level is to ensure that any newly compiled binary or malicious code within the cyber range is incompatible with any other regular computer system. A secure operating system should be used for this purpose; also, tools for compiling binaries should be provided. Since the environment focused on tool testing rather than real-machine simulation, one should keep all functions simple for researchers and newcomers to use.

─ *Vulnerable Binary Manager*

As shown in Figure 2, the purpose of the Vulnerable Binary Manager is to execute vulnerable binaries within the secure environment, and use input exploits against the vulnerable binary and analyze the effect of exploitations. This will limit all exploit activities within the Secure OS – thus protecting the host machine.

— *Vulnerable Binary*

A vulnerable binary is a file that contains purposefully designed vulnerabilities for automated tools to exploit. The complexity of the vulnerabilities should be simple and meaningful, which should be able to demonstrate one or more functions of the automated cybersecurity tool. The vulnerable binary should be able to execute within the Secure OS, but not in any generic operating system for our intended incompatibility purpose.

### 3.2.2 Generic Operating System (Generic OS)

Since automated cybersecurity could be conceptually developed for any platform, a generic operating system (Generic OS) can be selected as needed. In order to communicate with the cyber range, a vulnerability manager will be used for sending through exploits and receiving responses from the vulnerable binary manager.

— *Vulnerability Manager*

The vulnerability manager is a module that could take input exploit, communicate with vulnerable binary manager to how to exploit a certain vulnerable binary. This module should be able to be embedded or provide application programming interfaces(API) to other automated cybersecurity tools for fully automated experiments.

— *Automated Cybersecurity Tool*

Secure OS is incompatible with any Generic OS for security purpose, and may lack some critical data due to simplification, which could cause difficulty for cybersecurity tool development and less powerful, therefore, cybersecurity tools should be deployed within a Generic OS. This also allows the cybersecurity tool targeting the Secure OS to easily port or make modifications for other purposes.

— *Vulnerable Binary*

Like the vulnerable binary in Secure OS, a vulnerable binary in the Generic OS is a file that contains purposefully designed vulnerabilities for automated tools to exploit.

### 3.2.3    Sequence Diagram
Figure 3 below describes how a user could interact with our proposed cyber range.
Firstly, the users are required to setup up generic operating systems for both the virtualized cyber range and the environment for cyber attack tools. A virtual machine with a secure operating system should also be installed to the virtualized cyber range. Secondly, the user would start the vulnerable binary manager and vulnerability manager for the cyber attack tool to interact with the cyber range. The vulnerability manager could be embedded into cyber attack tool for easier managing. Thirdly, after the environmental setup, vulnerable binaries should be created, compiled and installed into the cyber range. Fourth, if a file-transfer function is not developed, vulnerable binaries should also be installed into the automated cyber attack tool for analyzing and exploiting. Finally, the user would use the automated cyber attack tool to exploit vulnerable binaries, generate exploits, then send exploits through the vulnerability manager to the cyber range for analyzing the effect of the exploit, and for collecting results.

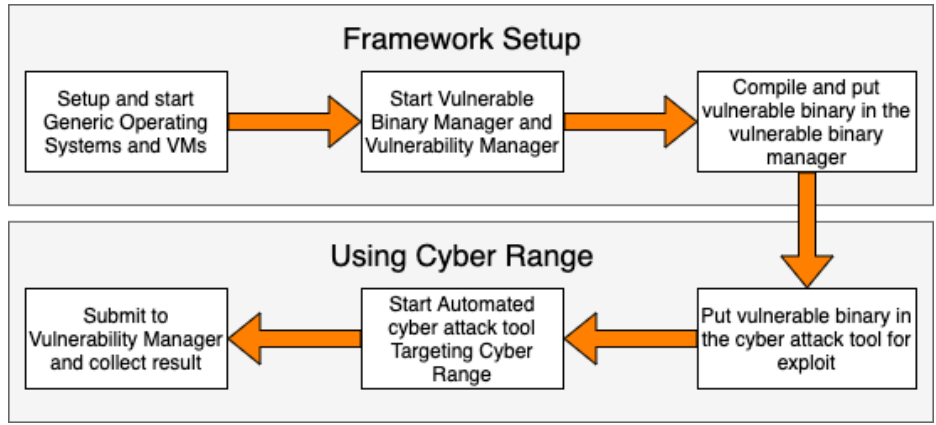

**Fig. 3.** Sequence Diagram

## 4  Case Study and Experiments

In this section, we will describe our implementation and experiments based on the scope defined earlier.

### 4.1  Framework Implementation

We implemented the experiment using the following equipment:

#### 4.1.1 Experiment Computer Hardware Specification

We intentionally implemented our experiments on a laptop to reduce the need for future researchers to procure specialized hardware. As such, we constrained our implementation around practical and commonly-accessible computing equipment to encourage reproducibility. As such, we implemented our experiment and recommend at least the following configuration:
- CPU: 4 processor cores
- Memory: 16 GB RAM
- Operating System: Windows 10/MacOS X 10.x or higher/Linux(kernel 4.0 or higher)

### 4.1.2 Virtual Machine Setting for Virtualized Cyber Range and Generic Operating System

Similarly, we implemented the virtualizes cyber ranged and Generic OS in the following environment:

- VMware Fusion 11.5.6
- CPU: 2 processor cores
- Memory: 8GB RAM

### 4.1.3 Virtualized Cyber Range Implementation

— *DECREE - Secure Operating System*

To fulfil the requirement of our proposed design, we used DECREE [34] created by DARPA for the CGC [16] as the base of the cyber range structure in the case studies. DECREE is a modified i386 Debian Wheezy with Linux kernel 3.13.2, with only seven modified system-calls and minimum non-determinism in system level [34]. As such, this increases the replayability of the system, which is highly useful for our research purpose.

DECREE has a built-in custom-made compiler and correlated toolsets, which can be used to compile the C source code into a unique executable format called DECREE Binary, which makes the DECREE binaries incompatible with a generic operating system.

For the case studies, a virtual-machines cluster developed by DARPA called virtual-competition [35] will be used. It contains five pre-set virtual machines that represent each significant part of the competition but does not have any abilities to do analysis[36]. Software for the virtual-competition will not be used in this study.

— *cb-server - Vulnerable Binary Manager*

cb-server [37] will be used as the vulnerable binary manager, which is responsible to take exploit inputs from the vulnerability manager, analyze the exploit and use it against the vulnerable binary within the cyber range.

— *Vulnerable Binary*

Legit_00003[38] is a simple C program created by LegitBS during DEFCON 24 in 2016. It was originally designed for human to human attack and defense CTF with CGC infrastructure. It contains a simple menu with three options. The function of legit_00003 is simply let the user input their name, print their name or exit. It has a basic stack overflow vulnerability which can be found with fuzzing or symbolic execution techniques.

### 4.1.4 Generic Operating System (Generic OS) Implementation

— *Operating System*

The operating system of choice for both case studies is the Ubuntu 18.04 LTS with 4.15.0-76-generic kernel.

— *Cb-Replay-Pov - Vulnerability Manager*

The program cb-replay-pov[37] will be used as the Vulnerability Manager, which is able to accept POV format file and communicate with the Vulnerable Binary Manager cb-server.

— *Legit_00003 - Vulnerable Binary*

The same binary for Virtualized Cyber Range (see Section 4.1.3) has been used. In this implementation, the exploit process remains in a generic Linux OS that has an angr development environment [39] installed. The vulnerable binary needs to be copied into the host machine in order to generate exploit, which shows that the architecture design of the proposed testing environment is feasible.

### 4.2 Implementation of Automated Cybersecurity Tool

To simulate the automated attack tool, a combination of a fuzzing tool and an automated exploitation engine have used to represent an automated attack tool. To make the combination using two tools, we have programmed a script application.

#### 4.2.1 Fuzzing Tool - Phuzzer

Phuzzer [40] is a Python wrapper for interacting with fuzzers created by the angr team; it has the American Fuzzy Lop (AFL) [41] as one of the core functions. Since the Cyber Grand Challenge considers a system crash as a vulnerability, this type of fuzzing methods for the simple program used in this case is an excellent choice. Therefore, in this case, Phuzzer is used to find the crash string from the vulnerable program.

#### 4.2.2 Automated Exploit Engine – rex

rex[42] is an automated exploit engine created by the team Shellphish, from the University of California Santa Barbara. It is a tool designed to generate exploit for DECREE binary and simple Linux Binaries. In our case study, rex is used to exploit targeted vulnerable programs, using the relevant crash string from the fuzzing tool. Note that due to dependency issues, rex is recommended to executed within an angr development environment.

#### 4.2.3 Script for Exploit

We wrote a script that combines phuzzer and rex to simulate the automated tool. The combined symbolic execution techniques and Phuzzer with AFL core are executed to find a string that could crash the legit_00003. After that, rex was executed to generate a Proof of Vulnerability (POV) (described in Section 4.3 later) format exploit. Figure 4 shows the core part of the code.

```
// Use phuzzer to crash legit_00003
python -m phuzzer -d 6 -C ./legit_00003

// Use rex to exploit legit_00003
crashFile = open(glob.glob("/dev/shm/work/fuzzer-master/crashes/id*")[0], 'rb')
crashes.add(crashFile.read())
...
workBinary = archr.targets.LocalTarget("legit_00003", target_os='cgc')
crash = rex.Crash(workBinary, crashes.pop())
crash.explore()
...
arsenal = crash.exploit()
arsenal.best_type1.dump_binary("legit_00003_type1.pov")
...
```

**Fig. 3.** Script for Automated Exploit an POV generation

### 4.3 Case Study 1 – Automated Exploit

To confirm the functions of our implemented design with DECREE and correlated-infrastructure, we carried out an experiment for receiving and checking exploits input for both manual and automated methods.

Since this case study builds on concepts from the DARPA CGC infrastructure, some information needs to be introduced in order to understand the procedures. For example, a Proof of Vulnerability (POV) is a type of XML or binary format file used to describe where is the vulnerabilities are and how to exploit the binary.

There is no binary execution within the host OS, nor any direct modification of the vulnerable binary. The POV data will be sent through Vulnerability Manager and received by Vulnerable Binary Manager inside of Virtualized Cyber Range. A faulty POV will result in Vulnerable Binary Manager return incorrect response, thus causing experiment failure.

#### 4.3.1 Assumptions for Automated Exploit

For the purpose of our experiments and for this research, we made a few assumptions:
— Since the goal of the study is to understand that if the designed framework works for automated attacks, we will not focus on the actual effects of the attacks.
— The focus of this study is to understand if the infrastructure functioned as intended. Hence, the actual contents of the POV will not be discussed.
— Manual attack processes will use pre-written C-format exploit created by the Legitbs which came along with the legit_00003 program.
— For our experiments, we assume that the vulnerability is known, so no vulnerability discovery processes will be made other than a simple fuzzing process.

#### 4.3.2 Testing Procedures

— *Running Vulnerable Binary Manager*

The application cb-server will be running in Secure OS as the Vulnerable Binary Manager, waiting for input to exploit Vulnerable Binary.

─ *Exploit Generation*

The provided exploit which came with legit_00003 will be used as a reference since it has been tested before and proved to be effective. In this study, this POV will be compiled and send to the Secure OS virtual machine with Vulnerable Binary Manager as a controlled group. The source code of legit_00003 and its C format POV can be found at [38]. The automated tool will generate a POV which represents an automated cybersecurity tool, subsequently uploaded to the Vulnerable Binary Manager, and the result will then be used for comparison.

─ *Exploit Vulnerable Binary*

During the process, data will be sent by the Generic OS virtual machine, through the program Vulnerability Manager. The host machine has no direct connection to the VM that is executing the binary file with SSH. On the Secure Operating System side, Vulnerable Binary Manager will handle the exploit and analyze the input.

### 4.4    Case Study 2 – Incompatibility of the DECREE System

In our second experiment, we aim to demonstrate the advantage and usefulness of isolating the virtualized cyber range by making the Virtualized Cyber Range environment incompatible with Generic OS. A simple application with an indefinite loop to represent a simple 'malware' was executed in our Virtualized Cyber Range and a Linux machine for comparing the test results.

To program the simple malware for our experiment, we used an uncompiled version of legit_00003 as the template. In order to make sure that there is no influence from legit_00003, all unnecessary files and DECREE system calls were removed. Figure 5 shows the source code of the simple application tested. As shown in the source code, it the program will indefinitely print out an integer with an increment of 1. Note that the header file 'printf.h' is from the package that came with the legit_00003, and a minimum version of 'printf' statement.

### 5    Result and Analysis

### 5.1 Case Study 1 – Result

In Case Study 1, we carried out exploiting Vulnerable Binary using an implemented automated cybersecurity tool to validate our implementation. The result of 'pov_1.pov' is the given POV from the legit_00003 creator, and the other one is the output from the automated attack tool. The provided POV from the legit_00003 creator was used as a reference to check the exploitation result of our attack tool.

There are two types of POVs[43]}:

```
#include "printf.h"

int main(void){
  int i = 0;
  while(1) {
    printf("%d\n", i);
    i++;
  }
  return 0;
}
```

**Fig. 5.** 'Malware' Source Code

- Type 1 POV requires to demonstrate control of instructor pointer and one of the additional for the vulnerable binary within DECREE, which apply to this case.
- Type 2 POV requires to read the contents of arbitrary memory location.

Figure 6 shows the result of both POVs, which consists of the following values.

```
**********************        **********************
POV Provided                   POV Automated Script
**********************        **********************
# pov_1.pov                    # legit_00003_type1.pov
# negotiation type: 1          # negotiation type: 1
# type 1 masks: 7f7f7f7f 7f7f7f7f  # type 1 masks: 7f7f7f7f 7f7f7f7f
# type 1 pov: 1e5d4c03 22392e5d 5  # type 1 pov: 00126a5e 4b6f5a78 5
```

**Fig. 6.** Result of both types of POVs

— *type 1 masks*
  • According to the documentation of POV [43], the combination of the IP mask and Register Mask form the exploitation output to Type 1 mask. IP mask and Register Mask are required values for the Vulnerable Binary Manager to exploit a vulnerable binary in DECREE. Based on the delivered IP mask value and Register Mask value, the attack point of vulnerable binary will be pointed out.
— *type 1 pov:*
  • Type 1 pov shows the IP value and Register value that can be used to exploit the target vulnerable DECREE binary. The successful obtaining of the type 1 pov value will allow the POV to assert that it can control the instruction pointer and eight general-purpose registers [43]. The instruction pointer is a kind of memory pointer for an executable file and the eight general-purpose registers are designed registers for DECREE to exploit binary files compiled for DECREE system. However, since the CGC infrastructure was not fully open-sourced, we are unable to analyze the value. That said, checking the acceptability of the framework design can be done sufficiently by comparing the generated data from our attack and the provided data from vulnerable binary designer.
  • The last digit of the *type 1 pov* value indicates which one out of eight valid general-purpose registers contains the required value [43]. Since the automated script received produced the same register as the provided one, it can be inferred and concluded that the value is correct.

### 5.2 Case Study 1 – Analysis

The functionality of our proposed design is our main focus for Experiment 1, which can be divided into two points: (1) communication and binary execution in two different environments with different executable formats, and (2) the performance of automated tool in such a special environment.

In this case study, the vulnerable binary manager cb-server is running within the secure operating system to handle the binary execution and exploit input. At the same time, the vulnerability manager is within a generic operating system that communicates with the vulnerable binary manager through the Transmission Control Protocol (TCP). This shows that, as long as a proper manager service is within each of the operating system, standard networking protocols could also make isolating the cyber range in the operating system level possible.

For the performance of automated tool, legit_00003, the vulnerable binary has a simple buffer overflow vulnerability leveraging the buffer size issue. The buffer overflow vulnerability enables attackers to take control of the computer system if they could control memory pointer to access some malicious code programmed by them. This type of buffer overflow attack is typical of many attacks experienced in the real-world today. The example POV contains a hard-coded input to fill out the buffer, and uses some techniques to exploit the target binary. Figure 7 shows the actual payload that exploits the legit_00003 from the POV provided.

```
...
senddata(1, "aaaaaaaaaaaaaaaaaaaaaaaaaaaaaaaa", 32);
senddata(1, "\xc4\xaa\xaa\xba", 4);
senddata(1, "junk", 4);
senddata(1, ®value, sizeof(value));   <-- "regvalue" and "ipvalue" was received
senddata(1, &ipvalue, sizeof(value));        from the server-side in previous part
...
```

**Fig. 7.** Payload from provided POV

Recalling Figure 6, it was shown that generated POVs by the automated attack tool were similar and provided POV except the value of 'type 1 pov' since they are representing specific memory addresses of the vulnerable binary, which should be different every time. Also, considering the 'cb-replay-pov' and 'cb-server' acceptance mechanism that only accept right POV report, we can confirm that the exploit generated by automated script works very similar to manually exploit provided. Thus, the performance of automated tools in the proposed environment works as intended.

### 5.3 Case Study 2 – Result

Figure 8 shows the shell output of the programmed simple 'malware' in the Virtualized Cyber Range virtual machine. As it was intended, the malware prints infinitely numbers of integers start from 1 with an increment of 1.

```
$ ./sampleMalware
1
2
3
...
```

**Fig. 8.** Shell output of the sample malware

To compare its execution result, the 'malware' was copied to two different Linux distributions (Ubuntu 18.04 and Debian 4.19.132-1). As shown in Figure 9, both operating systems failed to execute the sample malware with the same output.

```
bash: ./sampleMalware_00001: cannot execute binary file: Exec format error
```

**Fig. 9.** Shell output of the sample malware in a generic Linux System

### 5.4 Case Study 2 – Analysis

In Case Study 2, a sample 'malware' was successfully compiled into the DECREE binary format. As Figure 8 has shown, the 'malware' can be successfully executed within DECREE but, as expected, cannot be executed within a generic Linux system as showed in Figure 9.

The failure of executing the malware created for DECREE system in the generic operating system verified the program incompatibility between DECREE system and tested generic OS. This incompatibility confirms that automated cyber attack tools compiled for proposed cyber range are not able to propagate on to real-world computing systems linked to the cyber range in a corporate network (e.g. a Local Area Network (LAN)). The success of this case study also shows that the concept of using different operating systems in cyber range could provide a reasonable level of security while keeping the functionality of the cyber range. We acknowledge that there is also potential for further research to be conducted to verifying the security of our experiment through other techniques, such as formal verification.

Even though the process for creating the 'malware' was very simple with a vulnerable binary template, it was representative as the compile processes are very similar to the 'make' process of any generic C program. In order to create a DECREE binary without a template, a number of special modules are required as described in Figure 10.

Although it is possible to extract all tools and related modules from DECREE to create an individual compiling script, the vulnerable binary template is still our preference because of its convenience. This also applies to other secure operating systems with the same type of incompatibility design. It is worth noting that a cybersecurity tool may be required to do multiple compilations or script generation over time, which may result in serious performance degradation due to the constant access of the secure operating system. In such a situation, the tool should be developed in a generic operating system with all the required packages integrated.

```
binutils-cgc-i386         | Cross-binutils for CGC binaries
cgc-humint                | Human CGC Detector
cgc-network-appliance     | CGC Network Appliance
cgc-pov-xml2c             | CGC XML to C conversion for PoVs
cgc-release-documentation | Cyber Grand Challenge CGCOS documentation
cgc-sample-challenges     | CGC Sample Challenges
cgc-service-launcher      | CGC Service Launcher
cgc-virtual-competition   | Cyber Grand Challenge Virtual Competition
cgc2elf                   | Convert ELF binaries to CGCOS binaries
cgcef-verify              | Verify executables are in the proper CGC Executable Format
clang-cgc                 | LLVM/Clang for CGC
gdb-cgc                   | The GNU Debugger with CGC CB support
libcgc                    | CGC OS syscall library
libcgcdwarf               | CGC DWARF library
libcgcef0                 | CGC Executable Format Library
libpov                    | CGC POV library
linux-headers-3.13.2-cgc  | Linux kernel headers for 3.13.2-cgc on i386
linux-image-3.13.2-cgc    | Linux kernel, version 3.13.2-cgc
linux-source-3.13.2-cgc   | Linux kernel source for version 3.13.2-cgc
magic-cgc                 | CGC Magic
readcgcef                 | CGC readelf equivalent
services-cgc              | CGC Services
strace-cgc                | A system call tracer
```

**Fig. 10.** Packages required for compiling the sample malware

## 6  Concluding Remarks and Future Work

We proposed Pandora, a simple and secure cyber range framework to allow cybersecurity researchers to perform automated tool testing during the development process within a truly sandboxed environment. Our proposed design focus on the isolation of the test environments from real-world systems in order to address risk and malware propagation concerns for security testing of automated cyber attack tools. This addresses one of the development and testing needs of cyber autonomy, an emerging and popular cybersecurity research topic.

Our approach leverages developed open sourced software packages used during the DARPA Cyber Grand Challenge to segregate the secure experiment system from host infrastructure and to provide a communication channel between the safe system and generic system.

We demonstrated successful test activities of an autonomous hacking application and compared malicious program execution results on our secure system against prevalent systems. Our two experiments demonstrated that our design achieved automated binary exploitations while making sure that a compiled program for the secure system in the range is not executable in generic system outside the range.

For future work, as our current version works only in the Intel i386 architecture, we aim to extend the secure operating system's hardware compatibility – a requirement for expanding feasible test infrastructures. The improvement of automation for the secure cyber range is another requirement to enable increased machine learning model training. The implementation of secure, real-time file transfer function for Vulnerable

Binary Manager and Vulnerability Manager would be a critical step towards achieving fully automated cyber range.

Our work also has potential from an educational viewpoint. Since the cyber range works similar to a server that could take and evaluate exploit input, it might be possible to form a capture-the-flag (CTF) styled training or red-blue team training based on Pandora's design. The server for CTF training should be fully automated with a dynamic scoring system for tracking and analyzing the performance of the trainees. Finally, for students and newcomers with little experience working on security measures such as Address Space Layout Randomization and Structured Exception Handling Overwrite Protection [44], Pandora could enable trainees to train in a simple, security-oriented environments like DECREE, reducing configuration needs and freeing up time for actual training.

## 7  Access to Code and Artifacts

Our work builds on the work of many other open source projects and it continues beyond the time of publication. It is our intention to continue the open source nature of the project. All code created for this paper can be found in the following repository: https://cyber.uq.edu.au/repos/pandora-cyber-range

Other accompanying code and environments such as the angr development environment and legit_00003 can be found in the original authors' repositories.

## 8  Acknowledgements

Special thanks to DARPA who have created the DECREE system, virtual-competition and other infrastructure-related modules, which have been used during our experiments. Also, thanks to the team from Legitimate Business Syndicate who have created the vulnerable binary legit_00003, which was a critical part of the experiments. Finally, thanks to the angr team and Shellphish team, who have created and open-sourced their outstanding designs.